\documentclass[prl,twocolumn]{revtex4-1}
\usepackage{fullpage}
\usepackage{draftcopy}
\usepackage{graphicx}
\usepackage{verbatim}
\usepackage{color}
\usepackage{amsmath}
\usepackage{algpseudocode}

\begin{document}

\title{Beyond the Born-Oppenheimer approximation with quantum Monte Carlo}
\author{Norm M.~Tubman$^1$, Ilkka Kyl\"anp\"a\"a$^{3,1}$, Sharon Hammes-Schiffer$^2$, and David M.~Ceperley$^1$}
\affiliation{$^1$Department of Physics, University of Illinois, Urbana, Illinois 61801, USA\\
$^2$Department of Chemistry, University of Illinois, Urbana, Illinois 61801, USA\\
$^3$Department of Physics, Tampere University of Technology, P.O.~Box 692, FI-33101 Tampere, Finland}

\date{\today}
\begin{abstract}
In this work we develop tools that enable the study of non-adiabatic
effects with variational and diffusion Monte Carlo methods.  We
introduce a highly accurate wave function ansatz for electron-ion
systems that can involve a combination of both fixed and quantum ions.
We explicitly calculate the ground state energies of H$_{2}$, LiH,
H$_{2}$O and FHF$^{-}$ using fixed-node quantum Monte Carlo with wave function nodes that explicitly depend on the ion positions. The obtained energies implicitly include the effects
arising from quantum nuclei and electron-nucleus coupling. We compare our results to the best theoretical and experimental results available and find excellent agreement. \end{abstract}
\maketitle

\newpage

 From a computational perspective, treating quantum ions and electrons simultaneously appears to require significantly more effort than the pure electronic problem even though in principle similar techniques can be applied to both types of simulations.
One of the great successes in developing wave functions to go beyond
the Born-Oppenheimer approximation was the introduction of the
explicitly correlated gaussian (ECG) basis \cite{adam3,adam5}, which
allowed the prediction of ground state energies,
including non-adiabatic effects. Presently the ECG
method is limited to rather small system sizes and to only a few
quantum nuclei \cite{adam3,adam2}.  Other 
methods have also been introduced with the promise of treating
larger system sizes, such as nuclear-electronic orbital (NEO)
Hartree-Fock~\cite{shsnew1}, path integral Monte
Carlo \cite{Kylanpaa2012,Kylanpaa2011,Kylanpaa2010}, explicitly correlated NEO Hartree-Fock \cite{xchf3,xchf4,shs4} and
multi-component density functional theory \cite{shsnew2,shs1,shs2,shs3,shsnew3,shsnew4}.
However, there is not yet a clear path to simulating large system
sizes with high accuracy. 
 In this letter we develop tools for use in non-adiabatic quantum Monte Carlo (QMC) simulations
in order to include the effects arising from quantum nuclei and the
coupling between the electrons and the nuclei. We show that this
approach is competitive in accuracy with the ECG method, and
it can be extended to significantly larger system sizes.

 QMC methods have the capability to treat large
system sizes while maintaining highly accurate descriptions of the
electronic structure \cite{tubman1,foulkes1,grossman1,needs1,needs2,rothstein1,ceperley1,umrigar1,tubman2}. 
  An important component of these simulations is to generate a good starting wave function by determining the key variational degrees of freedom and then optimizing them in variational Monte Carlo (VMC).  Our approach assumes that the electron-electron correlations require the most variational degrees of freedom in the wave function, and to capture these correlations hundreds of parameters are introduced in the form of determinant coefficients.  We show that the electron-ion and ion-ion correlations can be treated to sufficiently high accuracy with a smaller set of variational parameters and that fixed-node diffusion Monte Carlo (FN-DMC) can be used to capture the remaining correlation energy.   
   
   The fixed-node approximation is used to simulate fermion wave functions with QMC.   It is based on the principle that a ground state energy can be determined exactly and efficiently if the exact nodes of the ground state wave function are known.  The nodes of a wave function are the regions of space in which the wave function is equal to zero.   This approximation is widely used because even when only approximate nodes are known, FN-DMC can be used to get a variational estimate of the ground state energy to high accuracy.  
   Only a few previous simulations of non-adiabatic Hamiltonians have been performed with FN-DMC, and they mostly have been limited to pure hydrogen systems~\cite{ceperley3,natoli1,natoli2,anderson1,nonada1,nonada2,bressanini1}. Some of these calculations are quite impressive as they involve the simulation of hundreds of quantum ions and electrons simultaneously.  Despite this success, there have only been a handful of FN-DMC calculations for small hydrogen molecules, and outside of this we are not aware of any other FN-DMC simulations for non-adiabatic Hamiltonians since the simulations on solid hydrogen~\cite{natoli1,natoli2}.

The adaptation of wave functions generated with typical quantum chemistry codes for use in non-adiabatic simulations is not straightforward. In particular, the single particle orbitals are dependent on the positions of the fixed nuclei, and it is not evident how to modify the single particle orbitals in a consistent manner.   
Several strategies can be implemented to use such wave functions in FN-DMC calculations, which are applicable to a wide range of problems having a combination of fixed and quantum nuclei.  
We consider three different wave function forms that are progressively more accurate as follows:
\begin{align}
\Psi(r,R) =& e^{J(r,R)}\phi(R)\sum_{i}\alpha^{*}_{i} D_{i}(r) \label{eqn:wfs1}\\
\Psi(r,R) =&e^{J(r,R)}\phi(R)\sum_{i}\alpha^{*}_{i} D_{i}(r,R^{*}) \label{eqn:wfs2}\\
\Psi(r,R) =& e^{J(r,R)}\phi(R)\sum_{i}\alpha^{}_{i} D_{i}(r,R), \label{eqn:wfs3}
\end{align}
where $r$ refers to the coordinates of all the electrons and $R$ to those of all the ions.  $J(r,R)$ is the Jastrow term which involves variational parameters that correlate the quantum particles and additionally  enforce cusp conditions in the wave function.  $\phi(R)$ is the nuclear part of the wave function. The final terms correspond to determinants $D$ and the corresponding coefficients $\alpha$.    The $*$ denotes how these terms are evaluated, as will be discussed. 

The nuclear part of the wave function is chosen to be a simple product of gaussian functions over each nucleus pair: 
\begin{align}
\phi(R) \propto \prod_{\substack{i \\ i<j}} e^{-a_{ij}\left(|R_{i}-R_{j}|-b_{ij}\right)^2}, 
\end{align}
where $a$ and $b$ are optimizeable parameters. In our calculations $a_{ij}$ has only a single optimized value $a$, and for $b_{ij}$ we use the Born-Oppenheimer equilibrium distance between the species involved.

The terms in these wave functions involve very specific calculations that are performed and optimized in both quantum chemistry codes and quantum Monte Carlo codes.  
The determinant terms, $\alpha_{i}^{*}D_{i}(r) $, $\alpha_{i}^{*}D_{i}(r,R^{*}) $, and $\alpha_{i}^{}D_{i}(r,R)$ differ based on how we optimize the determinant coefficients $\alpha$ and how we parameterize the evaluation of the determinants based on the ion coordinates $R$.   

The  wave function in Eq.~\eqref{eqn:wfs1} is the least accurate of the three wave functions and has a fixed determinant regardless of where the ions are.  The term $\alpha^{*}$ indicates that the determinant coefficients have been optimized at the equilibrium geometry.  
Both the ionic part of the wave function ($\phi$) and the Jastrow depend on the ion positions, which is important as the Jastrow maintains the cusps between all the quantum particles.  This form of the wave function has previously been used for large-scale simulations of metallic hydrogen~\cite{ceperley3,natoli1,natoli2}.    
The problem with this type of wave function is that the accuracy is limited by the electronic nodes, which do not depend on the ion positions.
This may be a good approximation for condensed matter systems, but in general  the determinant should depend on the ionic coordinates. 

The wave function in Eq.~\eqref{eqn:wfs2} fixes many of the problems of the previous wave function.  
The $\alpha^{*}$ indicates that the determinant part of the wave function is optimized for the equilibrium ion positions, as in the previous wave function, but the term $R^{*}$ signifies that the determinant  depends on the position of the ions through the basis set.  Basis sets in molecular calculations are generally constructed from local orbitals centered around the atoms.  In these calculations a single particle orbital is written as $\theta(r) = \sum_{ji}\gamma_{j}(r-R_{i})$, where \textit{i} is an index for an ionic center, and \textit{j} is an index for a basis set element.  
In this form, wave functions depending on the ion positions are straightforward to create and optimize,
but difficulties may arise with the possible directional dependence of the single body orbitals, such as in covalent bonds. This can be addressed with directionally dependent Jastrows, but we go further than this, as will be discussed.  This form of the wave function is similar to the wave function used in Ref.~\cite{ceperley3} for the molecular hydrogen phases.   They are not quite the same, however, as the electronic orbitals and ionic orbitals were centered around fixed positions and thus the electronic orbitals did not explicitly track the ion positions.   A few of the simulations did have the electrons track the centers of the hydrogen molecules as they changed position.

Eq.~\eqref{eqn:wfs3} represents what we expect to be the best wave function considered here, since it has explicit dependence on the ion positions for the single particle orbitals and the determinant coefficients. Essentially this amounts to recalculating a wave function from scratch each time the ion positions are changed.  This would significantly increase the computational cost of these simulations as well as cause many technical challenges. 

In this work we focus on the wave function in Eq.~\eqref{eqn:wfs2}, which is efficient and accurate, and captures the main physics of interest. To be more explicit, the wave function generation is done as follows with GAMESS \cite{gamess-1} and in a modified version of QMCPACK \cite{qmcpack1,qmcpack2}:
\begin{algorithmic}[1]
\State  Calculate a wave function in GAMESS for the fixed-ion
system of interest at the equilibrium geometry.
\State In case of multi-determinant wave functions, retain all the determinants with an initial coefficient larger than $\epsilon$, e.g., 0.0001. 
\State  Use an electron-ion cusp correction for the single particle orbitals.
\State  Optimize the $\alpha$ parameters and the Jastrow parameters simultaneously with the fixed-ion Hamiltonian.  We optimize one-body, two-body and three-body Jastrow terms for electron-electron and electron-ion coordinates.
\State  Optimize the ionic variational parameters in $\phi$ with the full electron-ion Hamiltonian.
\end{algorithmic}

\textit{Dragged Node approximation:}
It is useful to compare the different nodal structures in the wave functions given by Eqs.~\eqref{eqn:wfs2} and \eqref{eqn:wfs3}.  In Eq.~\eqref{eqn:wfs3}, the nodes are defined by the determinant that is calculated at each position in space, but in Eq.~\eqref{eqn:wfs2}, we use the determinant defined at the equilibrium geometry, and then drag those nodes around through the basis set dependence.  We call this the dragged node approximation, which is completely variational when used in VMC and FN-DMC.  
  In this work the ions obey Boltzmann statistics, which is exact for all the systems considered here. 
    A determinant can be introduced for ionic orbitals when the statistics of the ions is important.

 \textit{Single particle orbitals and relative coordinates:} The largest system we consider here is  FHF$^{-}$.  This is a linear molecule in which we fix the fluorine positions and treat the electrons and hydrogen nucleus quantum mechanically.  We use the form of Eq.~\eqref{eqn:wfs2}, without modification, as the fixed F ions localize the hydrogen ion between them.   Additionally we also consider LiH and H$_{2}$, which are rotationally symmetric systems for the ions.  Direct use of the fixed-ion wave functions causes an artificial increase in energy as several of the single particle orbitals are oriented along the initial bonding axis.  To attain the highest accuracy possible, the wave functions need to be modified in order to track the ions as they rotate around each other.   This can be addressed by 
   explicitly symmetrizing the electronic wave function with respect to the ionic rotations.

     There are a few ways of modifying a QMC code for this purpose, without making a new wave function call for the different rotational configurations of the ions.   
     One solution is to sample ionic configurations as normal, and then rotate the whole system such that the ions lie along the direction in which the single particle orbitals were generated.  We can use this procedure for the wave function in Eq.~\eqref{eqn:wfs2}, as our Jastrow and ionic orbitals are generated in relative coordinates.  For two atom systems we do this as follows:\\
  \begin{algorithmic}[1]
  \State Apply a shift $\mathbf{S}$, such that the first ion is shifted into its position of the fixed-ion calculation.
 \State Apply a rotation $\mathbf{U}$ to rotate the second ion along the original axis of the fixed-ion calculation.
 \State Shift and rotate all the electrons by $\mathbf{S}$ and $\mathbf{U}$.
 \State Evaluate the wave function amplitude, gradient and Laplacian in the new coordinates.
 \State Apply the inverse rotation and the inverse shift to the electron and ion coordinates using $\mathbf{U^{-1}}$ and $\mathbf{S^{-1}} $.
 \State Apply the inverse rotation to the gradient using $\mathbf{U^{-1}}$ .
  \end{algorithmic}
 Analytic gradients and Laplacians can be used for the electronic part of the wave function, but we use finite differences to calculate the ionic terms. 

\begin{table}[b]
\caption{Non-adiabatic ground state energies of H$_{2}$: symmetric Hartree-Fock (HF), symmetric full-CI (CI) and non-rotationally symmetric full-CI (CI-nr) refer to the trial wave function. The FN-DMC-full/CI results are our best results.  The term ''fixed'' indicates fixed nuclei results and ''full'' stands for quantum nuclei results. Energies are given in atomic units with one $\sigma$ error estimate in parenthesis.}
\begin{tabular*}{8cm}{@{\extracolsep{\fill}} cccc}
\hline\hline
 &HF& CI-nr &CI \\ 
\hline VMC-fixed& -1.1360(1) & &-1.1742(1) \\
 variance-fixed& 0.147 & &0.016 \\ 
 VMC-full& -1.1197(1) &-0.751(1)&-1.1617(1) \\ 
 variance-full& 0.15 &0.864 &0.021 \\
 FN-DMC-full& -1.1639(2) &-1.163(1)&-1.16401(5) \\ 
 variance-full& 0.122 &0.111 &0.021 \\ 
 \hline 
 \end{tabular*} 
 \begin{tabular*}{8cm}{@{\extracolsep{\fill}}ccc}
 Comparisons&Our Work& ECG \\ \hline &-1.16401(5) &
 -1.16402503084 \cite{adam7}\\
 \hline\hline
 \end{tabular*}
\label{tab:1}
\end{table}

\begin{table}[b]
\caption{Non-adiabatic ground state energies of LiH: See the caption of Table \ref{tab:1} and text for more details.}
\begin{tabular}{cccc}
\hline\hline
 &HF& CASSCF-nr & CASSCF \\
 \hline
  VMC-fixed& -8.06434  & & -8.0691(2)  \\
   variance-fixed& 0.035  & &0.013  \\
  VMC-full&  -8.0596(1) & -8.0$<$ &-8.0648(2)    \\
  variance-full& 0.036 &0.5$>$ &0.015  \\
  FN-DMC-full&  -8.0655(2) & -8.0646(3)&-8.06628(2)    \\
  variance-full&0.036&0.022 &0.015  \\
  \hline
 Comparisons&Our Work& ECG &Experiment \\
 \hline
&-8.06628(2) &-8.0664371 \cite{adam4}  & -8.0674 \cite{exp1,adam1} \\
\hline\hline
 \end{tabular}
\label{tab:2}
\end{table}

\textit{Results for  H$_{2}$: } The ground state for the hydrogen molecule is achieved exactly using
DMC.  This is because the electrons and ions have spin degrees of freedom such that the exact solution is a nodeless wave function. The best QMC results to date were simulated by Chen and Anderson to quite high accuracy~\cite{anderson1}.  In this case the quality of the trial wave function only
affects the convergence speed.  Therefore our interest in the H$_{2}$ molecule is to demonstrate properties of the variance of the energy and the accuracy that can be achieved with our wave function ansatz of Eq.~\eqref{eqn:wfs2}.   In Table \ref{tab:1} we report our results for the symmetric Hartree-Fock wave function (HF), the symmetric full-CI wave function (CI), and a non-rotationally symmetric full-CI wave function (CI-nr). The terms Hartree-Fock and CI refer to how we generate the starting single particle orbitals and determinant coefficients in GAMESS.   For the CI wave functions (and also the CASSCF wave functions in Table \ref{tab:2}), the determinant coefficients are reoptimized, along with a Jastrow and variational parameters for the ionic orbitals.  We use the Atomic Natural Orbital basis sets for all of our calculations \cite{roos1}.

The VMC energy for our best wave function, i.e.,~$-1.1617(1)$, is only 2~mHa higher than the exact result, which not only demonstrates the high quality of our wave function ansatz, but in comparison to the HF and CI-nr results, this shows the importance of both treating the electron correlations to high accuracy and rotationally symmetrizing the wave function.  

\textit{Results for LiH:}
Our results for LiH are demonstrative of what can be achieved with our techniques in comparison to other methods, and we report  an energy that is about $0.1$ mHa higher than the best ECG estimate for the non-adiabatic ground state energy of LiH, as shown in Table \ref{tab:2}.
Our final result is approximately 1 mHa higher than the experimental estimate \cite{exp1} following the analysis of  Scheu, Kinghorn and Adamowicz \cite{adam1}.  However, predictions made in a recent thermochemistry benchmarking study \cite{ZPEref} suggest better agreement, although the error bars are on the order of 0.5 mHa, which are too large to make any definite conclusions.  The most striking and systematic results come from strong convergence of the ECG results with basis set size \cite{adam4,adam3}.  These results appear to be converged well below 0.1 mHa,  and this would imply the fixed-node approximation for our wave function ansatz yields an error of only $0.1$ mHa.  

The origin of our fixed-node error may come directly from the fixed-ion optimization of the determinant coefficients.   The energy of our fixed-ion FN-DMC simulation is roughly $0.1$ mHa higher than the best ECG result with fixed ions we get -8.07045(2) and the ECG value is -8.070553(5).    
 Therefore improving the electronic nodes further within our current ansatz would likely increase our accuracy below $0.1$ mHa.  This is feasible within our approach, as we are not close to the limit of number of determinants we can optimize. 
  \\

\textit{Results for H$_{2}$O and FHF$^-$:}  We test two larger systems to demonstrate the scalability  of FN-DMC in treating more interesting systems.  We use only a Hartree-Fock starting point for the electronic part of the wave function and  make no attempt to calculate the best energies for H$_{2}$O or FHF$^-$, although it is possible that our energies might currently be the most accurate.    For the water molecule we treat all three ions as quantum particles. We are not symmetrizing the wave function as the hydrogen ions rotate around the oxygen ion.  This will constrain the full rotations of the hydrogen atoms and increase the kinetic energy slightly, as was discussed previously. For our previous results of LiH and H$_{2}$, this caused an error of $1$ mHa relative to our symmetric wave functions.

Using a single determinant H$_2$O wave function with the fixed-node approximation  gives an error of about $10$ mHa relative to the exact
value \cite{H2Oexact}. However, here we are mainly interested in the
energy difference between the fixed nuclei and the non-adiabatic
cases, which gives an estimate for the zero-point energy (including
the effects arising from the electron-nucleus coupling).  For the fixed nuclei case we obtain $-76.4221(6)$ Ha for the total
energy, and for the non-adiabatic case we obtain $-76.4012(14)$ Ha. This
yields an energy difference of $0.0209(20)$ Ha, which is in good
agreement with the best zero-point energy estimate of $0.0211$ Ha
\cite{ZPEref} for the water molecule. 
More detailed analysis of the water molecule and hydrogen
bonding is a subject of a later paper with a more accurate trial wave
function.

For the case of the bihalide ion FHF$^-$ we treat the proton as a
quantum ion, but we fix the fluorine nuclei, as they are significantly heavier.  
This also enables us to determine a potential energy surface in terms
of the distance between the fluorine atoms, including the coupling of
the electrons and the proton. Fitting our FN-DMC results at various
different fluorine distances to a Morse potential \cite{Morse29}, we
obtain for the internuclear F-F distance the value $R_0 = 2.3037(41)~$\AA,
which coincides with the experimental non-adiabatic estimate of
$2.304$~\AA~\cite{FHFexp,Sharon05}. The other Morse parameters are $D_0
= 200.3470(6)$ Ha and $\alpha = 0.0330(12)~$\AA$^{-1}$.

\textit{Discussion:}
 Chemically significant non-adiabatic applications are typically much larger than can currently be treated with ECG and full CI methods~\cite{big1,big2,ceperley3}.   The results in this work demonstrate that FN-DMC has great potential for simulating non-adiabatic systems, as it is both fast and accurate.   Additionally QMC can complement the ECG method in benchmarking, especially since the ECG method loses accuracy for larger molecules. The accuracies within the fixed-node approximation can be further increased to some extent for all the systems considered here, without incurring significant increases in the computational cost. For example the fixed-ion LiH results took under 100 cpu hours to calculate, and the non-adiabatic calculations were done in under 1000 cpu hours. 

 As for applications, we 
are able to consider system sizes well beyond the
largest ECG calculations, including systems with more than two quantum ions.  The bottleneck with FN-DMC calculations for non-adiabatic systems is less about computer time and more about devising forms for the wave functions~\cite{cederbaum1,worth1}, especially for highly non-adiabatic systems.  A strategic combination of the wave functions in Eqs.~\eqref{eqn:wfs2} and \eqref{eqn:wfs3} is likely to produce excellent VMC results for non-adiabatic systems, and FN-DMC is capable of treating such systems even without significantly improving the wave functions used in this work. Moreover, it is quite possible that other techniques such as NEO and ECG can be combined with our QMC approach to produce even more accurate results, and subtle details about our wave functions can be explored with techniques recently developed for use in QMC \cite{tubman3,tubman4,tubman5,blunt1}.

\textit{Conclusion:} In this letter we have demonstrated that FN-DMC is comparable to the best methods that can be applied to calculate non-adiabatic ground state energies, and we have shown highly accurate results on four different systems.  Our procedure takes advantage of standard quantum chemistry methods to create wave functions for these non-Born-Oppenheimer simulations. The techniques presented here create new opportunities for using FN-DMC in the studies of non-adiabatic molecular and condensed matter systems.

\textit{Acknowledgments}: We thank Andrew Sirjoosingh, Misha Pak, Sasha Soudackov, Jeongnim Kim, Jeremy McMinis, Hitesh Changlani, and Lucas Wagner for useful discussions. This work was supported by DOE DE-NA0001789, and we used the Extreme Science and Engineering Discovery Environment (XSEDE), which is supported by the National Science Foundation Grant No.~OCI-1053575. S.H.-S. acknowledges support from the National Science Foundation under grant CHE-1361293.

 \bibliography{refs}{}

\end{document}